\documentclass[iop,revtex4]{emulateapj}

\usepackage{journals_aas}   

\newcommand{\nolinkurl}{\url}

\shorttitle{OCARS catalog}
\shortauthors{Z. Malkin}

\newcommand{\gaia}{\textit{Gaia}}

\begin{document}

\title{A new version of the OCARS catalog of Optical Characteristics of Astrometric Radio Sources}
\author{Zinovy Malkin}
\affiliation{Pulkovo Observatory, St.~Petersburg 196140, Russia}
\affiliation{Kazan Federal University, Kazan 420000, Russia}
\email{zmalkin@zmalkin.com}

\begin{abstract}
A new version of the Optical Characteristics of Astrometric Radio Sources (OCARS) catalog is presented. This
compiled catalog includes radio sources observed in different VLBI programs and experiments that result in source
position determination, their redshift, and photometry in the visible and near-infrared bands. A cross-identification
table between the OCARS and other catalogs is also provided. The status of the catalog as of 2018 September 7 is
described in this paper. The OCARS catalog currently contains 6432 sources, of which 3895 have redshift data and
5479 have photometric data. Compared with the previous version, the current version has been enriched with
extended redshift and photometry information, and cross-identification with several catalogs in radio, optical,
infrared, ultraviolet, X-ray, and gamma-ray bands. The OCARS catalog is updated every few weeks on average to
incorporate new data that appear in the NASA/IPAC Extragalactic Database (NED), SIMBAD database, and in the
literature.
\end{abstract}



\section{Introduction}
\label{sect:intro}

The Optical Characteristics of Astrometric Radio Sources (OCARS) catalog contains the optical characteristics
of astrometric and geodetic radio sources that have been observed in different astrometric and geodetic
VLBI observing programs that result in accurate position determination.
The main goal of this work is to collect in one place the optical information related to astrometric radio sources
stored in different databases, catalogs, literature, etc.
When available, the physical type of object, redshift (z), and magnitudes in the visual and near-infrared (NIR)
bands are included.
Since the publication of the first version of the OCARS catalog in 2007 December \citep{Malkin2008f}, the catalog
has been continuously developed by increasing the number of sources, adding new data, and refining existing data.
The current state of the OCARS catalog as of 2018 September 7 is described in this paper.

The initial goal of the OCARS project was to revise outdated redshift information that was included as a supplementary
data for the first International Celestial Reference Frame (ICRF) release \citep{Ma1998} because
the redshifts are relevant to cosmology and other scientific studies \citep{Titov2009b}.
The first version of the OCARS catalog was created as a supplemental data for the second ICRF release,
ICRF2 \citep{Fey2015}.
A more detailed description of the first version of the catalog is given in \citet{Titov2009b}.
Compared with the previous version, the current version has new content, such as extended redshift information,
{\gaia} photometry, and cross-identification table with several catalogs in different frequency bands
from the infrared (IR) to gamma-ray.

The primary sources of optical characteristics are the 
NASA/IPAC Extragalactic Database\footnote{http://ned.ipac.caltech.edu} (NED) and
SIMBAD\footnote{http://simbad.u-strasbg.fr/simbad/} database managed by the Centre de Donnees astronomiques de Strasbourg (CDS).
These services complement each other well.
Another important survey used for OCARS is the Sloan Digital Sky Survey\footnote{http://www.sdss.org} (SDSS)
\citep{SDSS14}, from which many redshift and photometric data have been obtained.
In earlier OCARS versions, useful information was also obtained from
Australia Telescope 20~GHz (AT20G) Survey \citep{Mahony2011},
Catalog of Faraday Rotation Measures and Redshifts for Extragalactic Radio Sources (RM-Redshift) \citep{Hammond2012m},
DEEP2 Galaxy Redshift Survey \citep{Newman2013}, 
Micro-Arcsecond Scintillation-Induced Variability (MASIV) Survey \citep{Pursimo2013},
Large Quasar Astrometric Catalogue (LQAC) \citep{Souchay2015}, and Milliquas \citep{Flesch2015} catalogs.
The latter is regularly maintained and is currently used for OCARS update.

To incorporate all possible sources of information and provide timely OCARS update with new data, all available
publications in journals and arXiv\footnote{https://www.arxiv.org} are permanently observed.
The OCARS catalog is updated every few weeks on average.
A total of 117 updates have been issued to date.

In addition to incorporating all the data available from the literature, targeted observing programs were also initiated.
The first observing program that aimed to determine redshifts of astrometric radio sources started at
the Pulkovo Observatory in 2008.
Several time slots for observations on the 6-m BTA telescope of the Special Astrophysical Observatory (SAO)
of the Russian Academy of Sciences were allocated for this program in 2008--2011.
For various reasons, primarily unfavorable weather conditions, these observations resulted in new redshift
determinations for only 7 objects \citep{Maslennikov2010}.
Observations with other telescopes, such as New Technology Telescope (NTT) in Chile,
Gemini observatory's telescopes in Hawaii and Chile, and Nordic Optical Telescope (NOT) in the Canary Islands,
that were later organized by Oleg Titov \citep{Titov2011b,Titov2013b} were much more effective and resulted
in determination of redshifts for 270 OCARS sources.
It is especially important that large number of these sources are located in the southern hemisphere where
the fraction of radio sources with known optical characteristics is lower than that in the northern
hemisphere (see Table~\ref{tab:stat}).

Work on a new ICRF version, ICRF3, began in 2012 \citep{Jacobs2014} and was finished in August 2018
(Charlot et al., 2018, in preparation).
The ICRF3 has been endorsed by the XXXth IAU General Assembly, Vienna, 20--31 August  2018.
A few months earlier, in April 2018, the second version of the {\gaia} Celestial Reference Frame, \gaia-CRF2,
was released \citep{Lindegren2018,Mignard2018}.
Comparison of these catalogs is important for the mutual aligning of the radio and optical celestial reference frames.
The quality of this comparison depends critically on the number of common sources between ICRF and \gaia-CRF
catalogs.
While {\gaia} provides positions of all the technically available objects, including extragalactic radio sources,
with compatible accuracy, the accuracy of the VLBI-derived positions is not uniform over the sky.
Only relatively small number of radio sources have the accuracy of the VLBI-derived positions better than 0.1--0.2 mas
because achieving such overall level of accuracy requires to collect approximately 200--500 observations for each
source \cite{Malkin2018}, which is practically impossible as astrometric VLBI observing resources are severely limited.

On the other hand, many radio sources included in the astrometric VLBI catalogs are too faint in optical to be
observed with {\gaia}.
Data of Table~\ref{tab:cross_identification} below shows that $\sim 30\%$ of OCARS sources are not matched to {\gaia} DR2.
\citet{Petrov2018} found that only 9081 of 15155 radio sources from the rfc2018b catalog ($\sim60\%$) are matched to {\gaia} DR2.
Therefore, OCARS helps in selecting radio sources that are sufficiently bright in optical for extensive observations
with VLBI to improve their radio positions and thus enforce the link between the radio and optical frames, as made
during preparation of ICRF3 \citep{Bourda2008,Bourda2010,LeBail2016}.

Another useful function of the OCARS catalog is establishing correspondence between objects in various astrometric
and astrophysical catalogs, which is not always obvious and unambiguous.
To address this task, a third file with a table of cross-identifications of objects in various catalogs and
databases was added to the OCARS catalog.

The paper is organized as follows.
Section~\ref{sect:description} presents a general description of the OCARS catalog,
Section~\ref{sect:photometry} provides a detailed explanation of the photometric data included in the OCARS,
Section~\ref{sect:crossid} describes the OCARS cross-identification table,
and Section~\ref{sect:conclusion} concludes the paper.


\section{General catalog description}
\label{sect:description}

The current version of the OCARS catalog\footnote{http://www.gaoran.ru/english/as/ac\_vlbi/\#OCARS} includes the following three files:

\verb"ocars.txt" --- the main catalog file that contains for each radio source B1950 and J2000 names, equatorial
and Galactic coordinates, and, when available, source physical type, redshift, approximate magnitude, and notes;

\verb"ocars_m.txt" --- contains photometric data in 14 frequency bands;

\verb"ocars_n.txt" --- contains cross-identification table of the OCARS with other catalogs.

The list of radio sources included in the OCARS catalog, along with their positions is formed from the
ICRF3\footnote{http://iers.obspm.fr/icrs-pc/newwww/icrf/} source list (P.~Charlot et al., in preparation)
containing 4588 sources (71.3\% of the OCARS list), and sources available from publications, such as
\citet{Momjian2004,Frey2005,Kunert-Bajraszewska2006,Frey2008,Bourda2010,Frey2010,Immer2011,Frey2011,Petrov2011a,%
Petrov2011b,Petrov2011c,Petrov2011d,Petrov2012a,Yang2012,Petrov2012b,Frey2013,Petrov2013,Cao2014,Schinzel2015,%
Gabanyi2015,Coppejans2016,Cao2017,Shu2017,Li2018}.
A catalog code indicating the original reference for the source position is given for each OCARS source
in the \verb"ocars.txt" file.

Each OCARS source has 8-character B1950 and 10-character J2000 names.
When a new source is included in OCARS, the B1950 name recommended by the IVS (International VLBI Service for
Geodesy and Astrometry, \citet{Nothnagel2017}) Source Name Translation
Table\footnote{https://vlbi.gsfc.nasa.gov/output/IVS\_SrcNamesTable.txt} is used if the source is present
in this table.
Otherwise, the name from the publication is used.
This rule provides a compliance with astrometric VLBI works coordinated by the IVS that traditionally
use B1950 8-character name as the primary source identifier.
The J2000 source name is derived from the source position.

Three columns of the \verb"ocars.txt" file contain redshift measurements found in the NED, SIMBAD, and SDSS.
Two latter values are always provided as returned by SIMBAD and SDSS in response to inquiries related to the source.
The first redshift column can contain either an NED value or a recent value taken from the literature.
In the case of using a non-NED estimate, the corresponding number is marked and explained in the notes.
Part of the redshift data, especially for BL Lac-type objects, is unreliable or ambiguous;
for some sources, only photometric redshifts, redshift estimates, or even redshift lower limits are known.
For such objects, special marks are used and comments are provided.

Approximate magnitudes listed in the \verb"ocars.txt" are determined as follows.
If photometric information for a source is available in the file \verb"ocars_m.txt"
(see Section~\ref{sect:photometry} for detailed description of the OCARS photometric data),
then the magnitude is taken from that file in the following order of preference: $RVGrgBiIzuUJHK$.
Otherwise, a tentative magnitude from NED is used without the band identification.
Comparison with SIMBAD shows that the NED tentative magnitude usually corresponds to the $B$ or $V$ bands.

Table~\ref{tab:stat} presents the overall statistics for the OCARS catalog.
This statistics shows that as many as 66.0\% of OCARS sources have both redshift and photometric data.
For ICRF3 sources, full optical information is known for 85.1\% of sources.
The fraction of these sources is a bit smaller than the fraction of sources with redshift data.
The fraction of sources with photometric information is much larger,~-- 85.2\% for all sources,
and 96.7\% for ICRF3 sources.

\begin{figure}
\centering
\includegraphics[clip,width=\hsize]{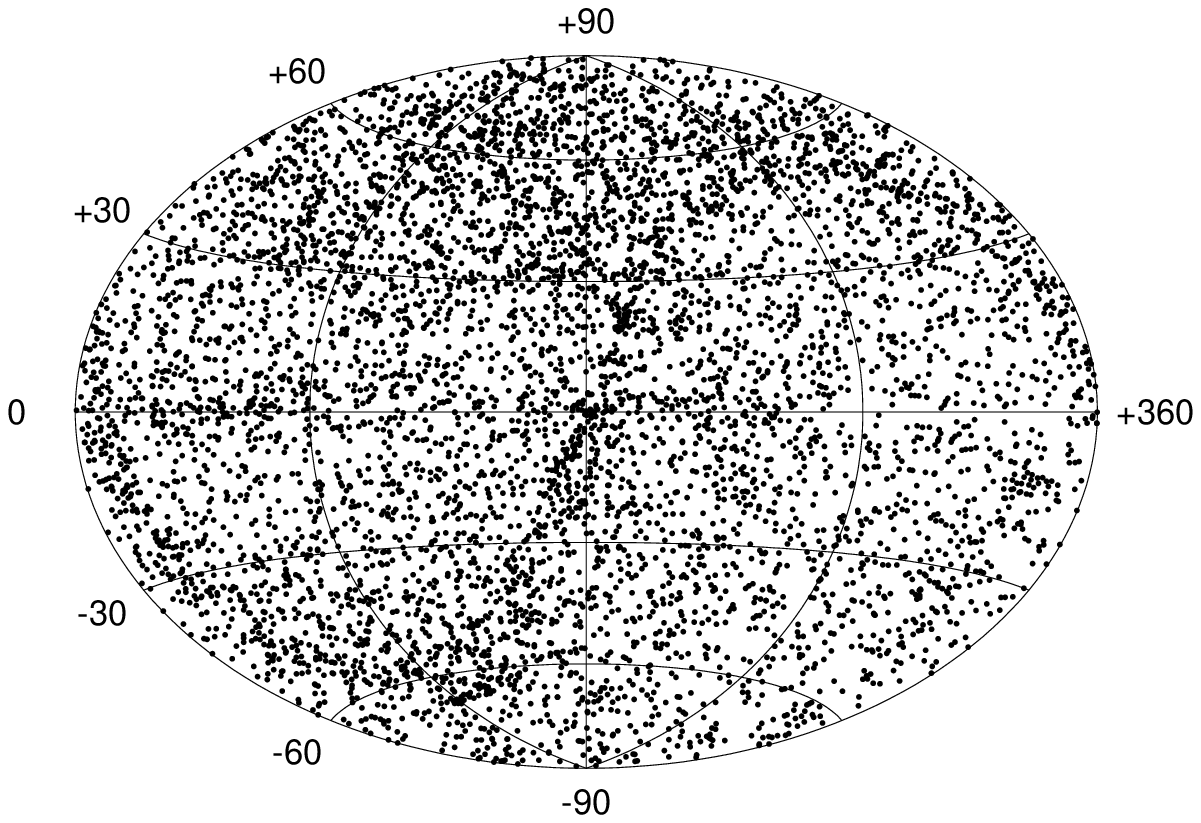}
\includegraphics[clip,width=\hsize]{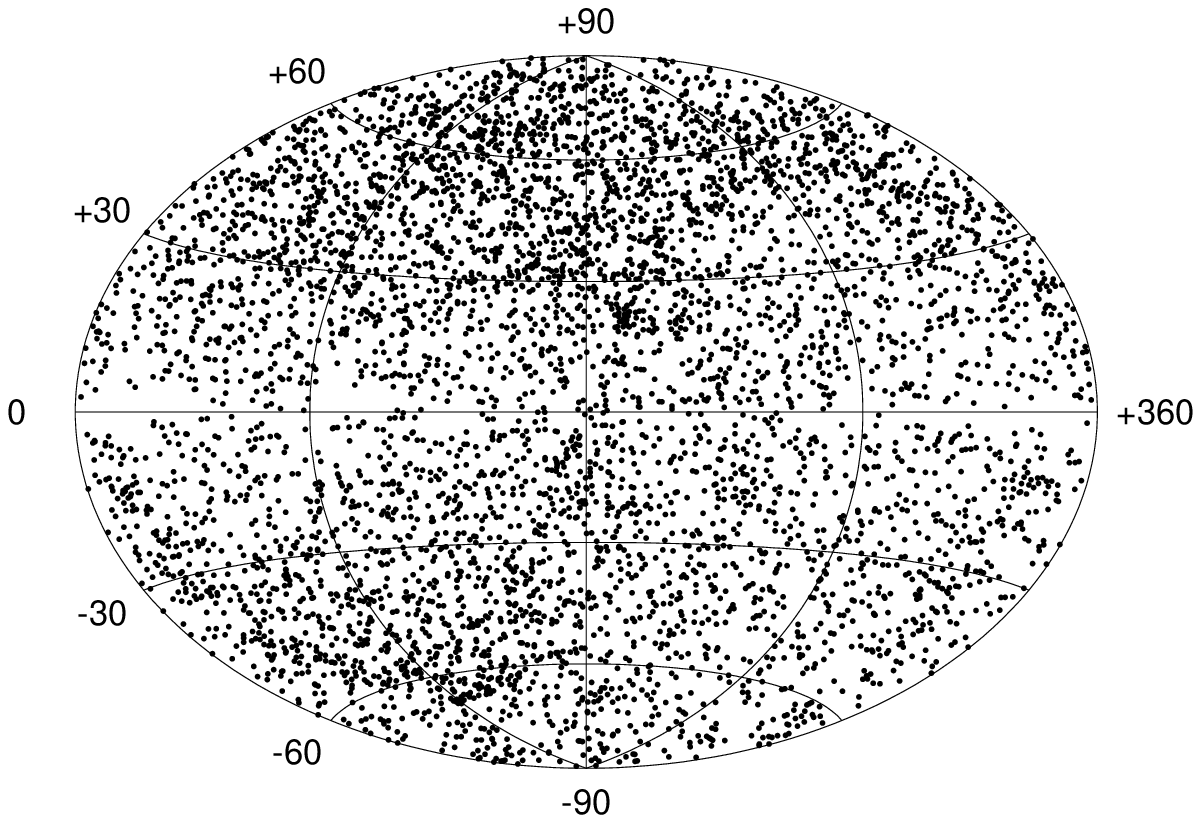}
\includegraphics[clip,width=\hsize]{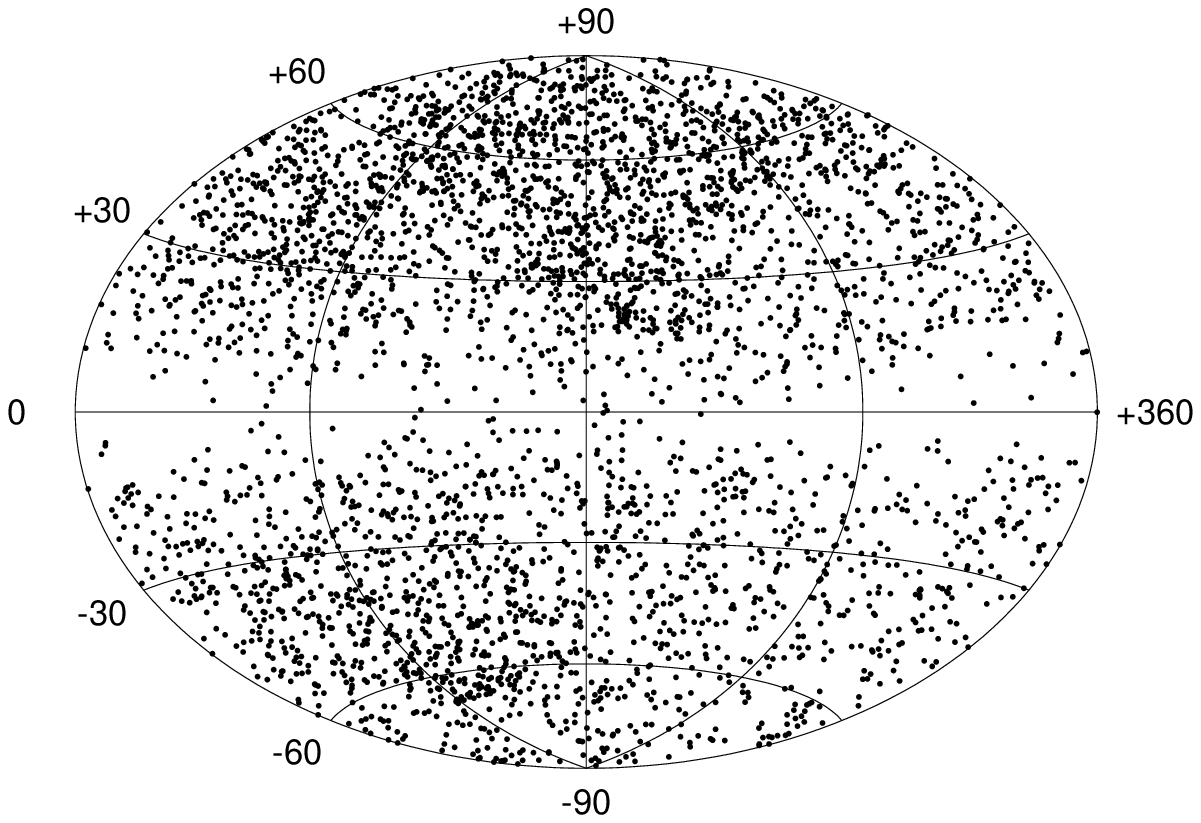}
\caption{Sky distribution of all OCARS sources (top) and sources for which photometry (middle) and redshift (bottom)
  data are present in OCARS (Galactic system, Hammer projection).}
\label{fig:sky}
\end{figure}

The distribution of the OCARS sources over the sky is shown in Fig.~\ref{fig:sky}.
It is not uniform for both all the sources and sources having photometric and redshift data,
with a deficit in the southern hemisphere.
Additionally, densification of sources near the ecliptic coming from \cite{Shu2017} and dispersion of sources
near the Galactic plane are observed.
A jump in the source distribution at $\delta \approx 40^{\circ}$ corresponds to the southern boundary
of the declination zone available for the Very Long Baseline Array (VLBA) that provided observations of most
astrometric radio sources.
Many sources were observed in the framework of IVS observing programs, which also mostly include sources
in the northern and equatorial declination zones.
Despite significant efforts to extend observation facilities and organize new VLBI observing programs
in the south in recent years \citep{Petrov2011a,Lovell2013,Basu2016,deWitt2016,Plank2017}, a relative
number of observations of southern sources remains approximately the same with time \citep{Malkin2015}.

\begin{table}
\centering
\caption{OCARS statistics for all sources and for ICRF3 sources. The statistics, except the source type, are given for all OCARS sources and for $45^{\circ}$ declination zones.}
\label{tab:stat}
\begin{tabular}{lrr}
\hline
& All~ & ICRF3 \\
\hline
sources & 6432 & 4588 \\
\qquad $+45^{\circ}$...$+90^{\circ}$  & 1120 & ~802 \\
\qquad ~~~~$0^{\circ}$...$+45^{\circ}$ & 2734 & 1864 \\
\qquad $-45^{\circ}$...$0^{\circ}$ & 2045 & 1647 \\
\qquad $-90^{\circ}$...$-45^{\circ}$  & ~533 & ~275 \\[1ex]
with known type           & 4079 & 3181 \\
\qquad  quasar            & 2473 & 1996 \\
\qquad  BL Lac            & ~642 & ~473 \\
\qquad  Seyfert           & ~274 & ~226 \\
\qquad  radio galaxy      & ~521 & ~359 \\[1ex]
with redshift ($z$) info  & 3895 & 3034 \\
\qquad $+45^{\circ}$...$+90^{\circ}$  & ~689 & ~514 \\
\qquad ~~~~$0^{\circ}$...$+45^{\circ}$ & 1886 & 1409 \\
\qquad $-45^{\circ}$...$0^{\circ}$ & 1070 & ~958 \\
\qquad $-90^{\circ}$...$-45^{\circ}$  & ~250 & ~153 \\[1ex]
with photometry info & 5479  & 4102 \\
\qquad $+45^{\circ}$...$+90^{\circ}$  & ~943 & ~705 \\
\qquad ~~~~$0^{\circ}$...$+45^{\circ}$ & 2342 & 1699 \\
\qquad $-45^{\circ}$...$0^{\circ}$ & 1713 & 1442 \\
\qquad $-90^{\circ}$...$-45^{\circ}$  & ~481 & ~256 \\[1ex]
with both $z$ and photometry & 3883 & 3026 \\
\qquad $+45^{\circ}$...$+90^{\circ}$  & ~684 & ~511 \\
\qquad ~~~~$0^{\circ}$...$+45^{\circ}$ & 1882 & 1406 \\
\qquad $-45^{\circ}$...$0^{\circ}$ & 1067 & ~956 \\
\qquad $-90^{\circ}$...$-45^{\circ}$  & ~250 & ~153 \\[1ex]
\hline
\end{tabular}
\end{table}

\begin{figure}
\centering
\includegraphics[clip,width=\hsize]{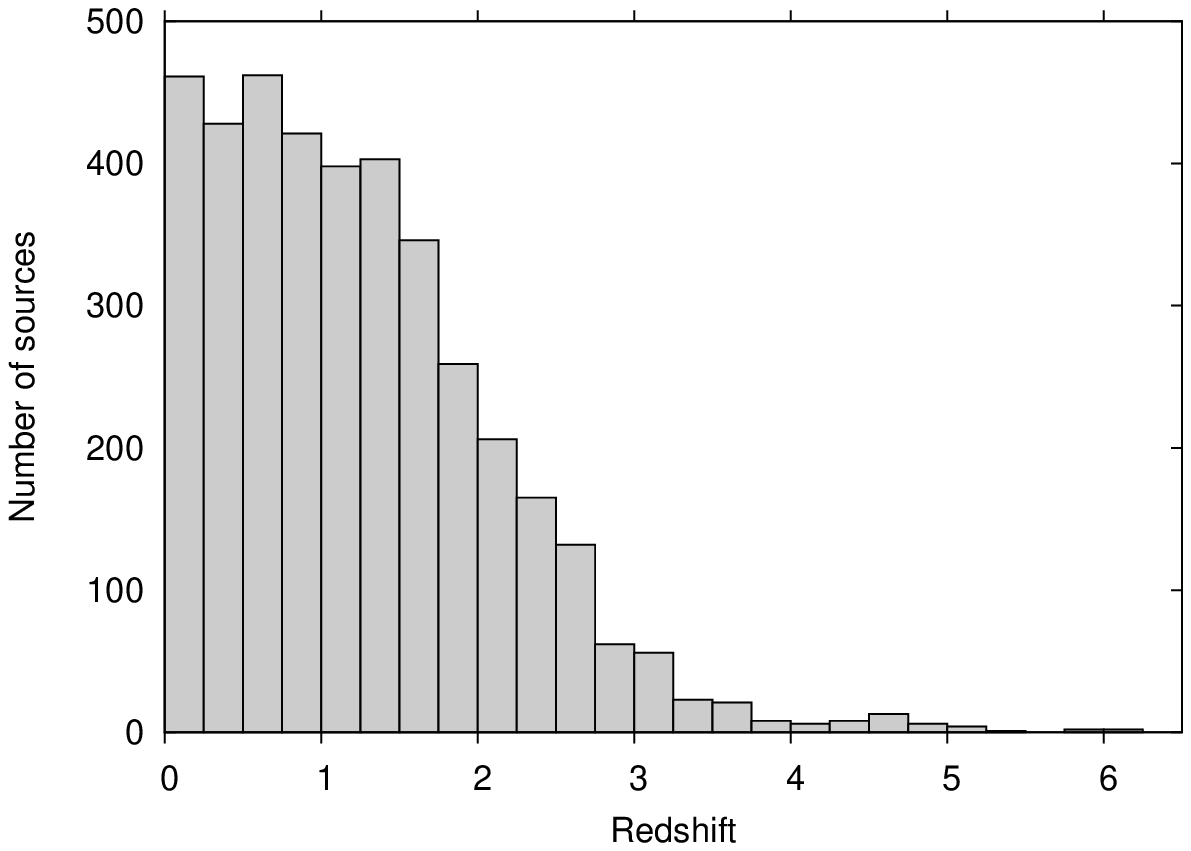}
\includegraphics[clip,width=\hsize]{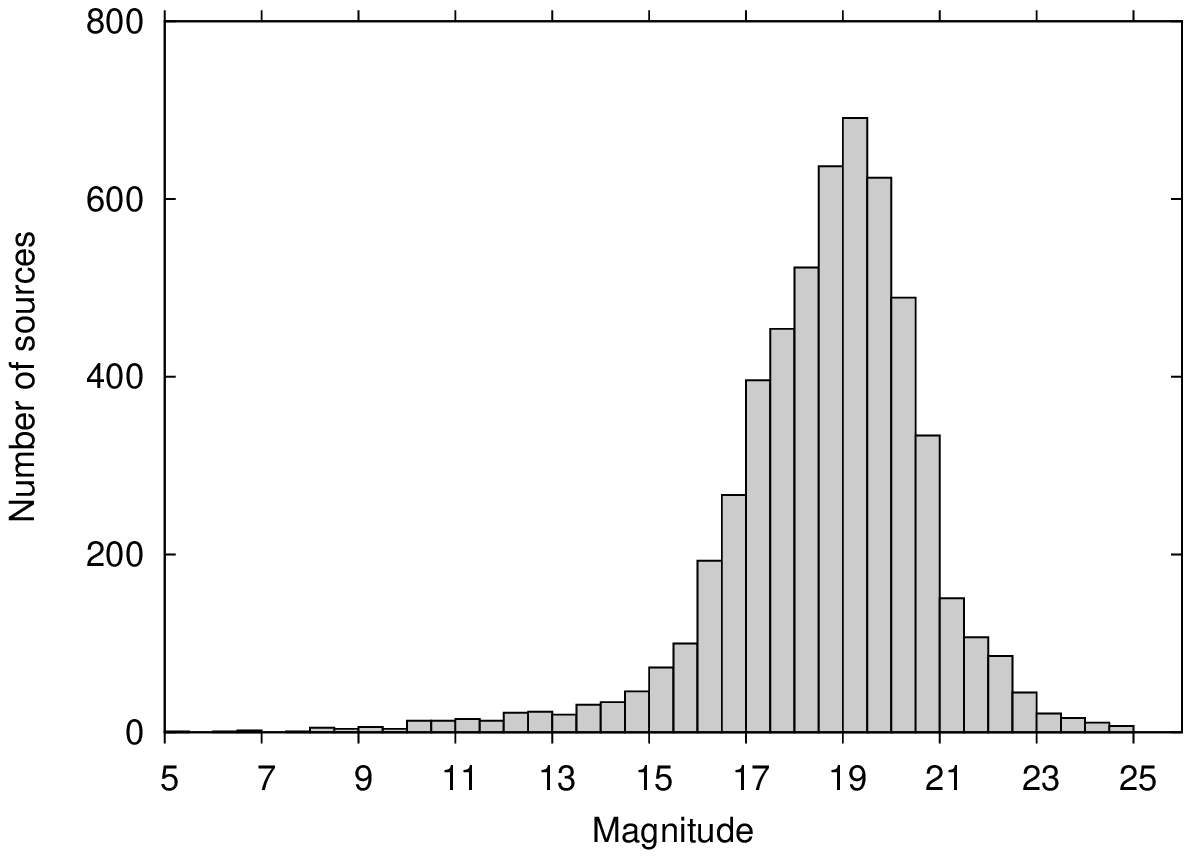}
\caption{Redshift (top) and magnitude (bottom) distribution for OCARS sources.}
\label{fig:z_mag_n}
\end{figure}

The redshift and magnitude distributions of the OCARS sources are presented in Fig.~\ref{fig:z_mag_n}.
The mode of the magnitude distribution of the OCARS sources is $\sim$19$^m$.
The optically weakest sources have magnitudes of $\sim 25^m$, however, this result refers only to optically
identified sources with measured optical magnitude.
The OCARS catalog contains sources with redshift up to $z=6.2$; 43 sources have $z \ge 4$, and 154 sources
have $z \ge 3$.

\section{Photometric data}
\label{sect:photometry}

Detailed photometric data for OCARS sources is provided in file \verb"ocars_m.txt".
This file contains magnitudes for the 14 bands similar to SIMBAD, namely, $uUBgVrRiIzJHKG$.
Data for close bands, such as $u/U$, $r/R$, and $i/I$, are not merged to preserve all the information
from the input sources, enable control of the OCARS data, and identify potentially unreliable data.
Only a few measurements at the $Z$ band have been merged with those at the $z$ band.

Photometry information is primarily taken from the NED and SIMBAD.
When flux density is given in NED instead of magnitude, the {\it Spitzer} magnitude--flux density
converter\footnote{http://ssc.spitzer.caltech.edu/warmmission/propkit/pet/magtojy/} is used to compute magnitude.

Statistics for the OCARS photometric data are presented in Table~\ref{tab:photometry_statistics}.
The first row of this table presents the number of sources that have photometric data in a given band.
The largest number of brightness measurements is available for the $R$, $B$, and $G$ bands.
The second row of the table provides the averaged color indices computed as the magnitude in a given band minus
$R$ magnitude.

A total of 444 OCARS sources have photometric data in only one band, and 189 sources have photometric data
in only two bands.
Only 23 sources have photometric data in all the 14 bands.
Only {\gaia} photometry is available for 273 sources.

For 29 sources, only  NIR photometry is available.
For these sources, the averaged color index can be used to estimate the magnitude in a optical band of interest
with a reasonable level of confidence.
Therefore, these sources can also be used to preliminary select objects suitable for the link between radio and
optical frames.
However, it should be kept in mind that objects bright in NIR are mostly extended galaxies whose positions cannot
be determined with {\gaia} \citep{deSouza2014}.

\begin{table*}
\centering
\caption{Photometry statistics and averaged colors.}
\label{tab:photometry_statistics}
\begin{tabular}{lcccccccccccccc}
\hline
Band            & $u$  & $U$  & $B$  & $g$  & $V$  & $r$  & $R$  & $i$  & $I$  & $z$  & $J$  & $H$  & $K$  & $G$  \\
\hline
N               & 1884 & ~357 & 4638 & 3322 & 2902 & 2966 & 4633 & 1915 & 2056 & 1883 & 2227 & 1866 & 2270 & 4579 \\
Color (band--R) &  1.0 &  0.2 &  0.8 &  0.7 &  0.5 &  0.2 &  --- &  0.1 & --0.5& --0.1& --1.4& --2.0& --2.8& 0.4  \\
\hline
\end{tabular}
\end{table*}

Some comments are necessary to convey a better understanding of the photometric data present in the OCARS catalog.
The emission of the extragalactic radio sources that define the optical and radio celestial reference frames
is primarily generated by active galactic nuclei (AGNs), which are often characterized by substantial variability.
Monitoring of the optical brightness of astrometric extragalactic objects is being conducted by
\citet{Taris2013,Taris2016,Taris2018}.
The results of this monitoring show that the optical brightness of astrometric radio sources usually varies
within several tenths of a magnitude, although the spread of the brightness variations often exceeds
a magnitude and can reach up to three magnitudes, as in the case of AGN 0716+714.
Of course, the latter case is not typical.
Analysis of literature data for the OCARS sources in the NED database also shows that photometric brightness
measurements for a given object within a given band can differ by more than one magnitude.

Because of variability, the optical magnitudes given in the OCARS catalog can be considered as indicative
and can differ from the instant magnitude value at a specific epoch.
However, their accuracy is mostly sufficient for planning of observations.
A complete description of the optical brightness would include the peak-to-peak interval of magnitudes
that can be observed for a given object, but such information is available for only a small number of sources.
The current version of \verb"ocars_m.txt" provides the average magnitude when several different measurements for
a band are present in the NED or literature.

Another source of uncertainty in the magnitudes is ambiguity in the SDSS data from which most $ugriz$ magnitudes
are taken.
The SDSS magnitudes are determined using several models\footnote{http://www.sdss.org/dr14/algorithms/magnitudes/}.
The differences among the values obtained for different models can reach a magnitude or more.
In accordance with the recommendations of the authors of the SDSS, Composite Model Magnitudes, \verb"cModelMag",
are used in the OCARS catalog.

Finally, bright stars are located in the sky near several extragalactic radio sources, which could be confused with
the OCARS objects in optical observations.
All known such cases are commented in the notes.


\section{Cross-identification table}
\label{sect:crossid}

The cross-identification table of OCARS sources with objects in other catalogs is provided in a separate file,
\verb"ocars_n.txt".
It is primarily based on the alias lists provided by NED and SIMBAD.
For some sources, OCARS includes new associations.
The latter is suggested if the following two conditions are fulfilled: the distance between the objects is within
the positional error ellipse, and no other object in NED and SIMBAD is found near the OCARS source that would
compromise a cross-identification suggestion.
In a few opposite cases, wrong identifications in NED or SIMBAD are commented in the notes.

Cross-identification of the OCARS sources with objects in large catalogs may lead to false
identifications in a case when two or more objects in an outer catalog that can be associated with the OCARS
source are present within the search radius.
In some cases, two or more OCARS entries have the same identifier in other catalogs.
As a rule, this situation occurs for objects with a complex structure, for which two or more centers of radio
emission are resolved with VLBI.

To eliminate the number of misidentifications, several measures are taken.
For the {\gaia} catalog, the program that searches for object matching outputs all the {\gaia} objects that fall
within the search radius.
Then, the nearest object is accepted for the OCARS cross-identification table.
In fact, there are only a few such cases.
During comparison of OCARS with LQAC and Milliquas, when several objects from an outer catalog are found within
the search radius, the object with redshift in that catalog is selected.
For radio catalogs, the matching object in crowded areas is selected by manual inspection of NED entries.
In case of doubts, a note is provided in OCARS.

\begin{figure}
\centering
\includegraphics[clip,width=\hsize]{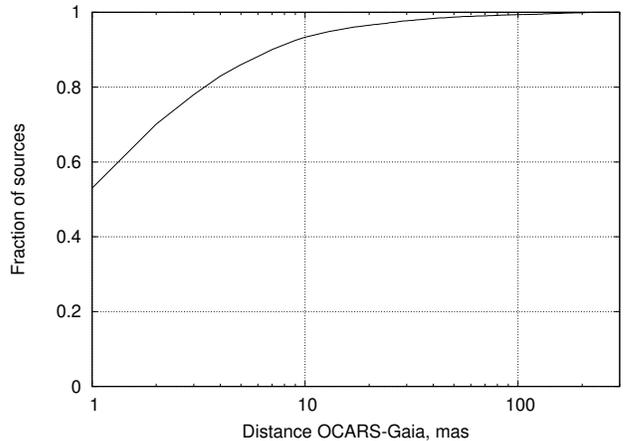}
\caption{Fraction of the OCARS sources matched to the {\gaia} DR2 as a function of the OCARS--{\gaia} arc length.}
\label{fig:cumulative_fraction}
\end{figure}

Cross-matching with {\gaia} was performed using a search radius of 0.15~arcsec for the ICRF3 catalog,
and 0.25~arcsec for other VLBI-derived positions.
The search radius for cross-matching ICRF3 sources with {\gaia} was chosen to be somewhat larger that 0.1~arcsec
accepted in \cite{Lindegren2018} to positionally match Gaia with VLBI sources.
Increasing the search radius from 0.1 to 0.15~arcsec added three more ICRF3 sources, from 3413 to 3416.
The fraction of the OCARS--{\gaia} counterparts as a function of the angular distance between OCARS and {\gaia} 
is shown in Fig~\ref{fig:cumulative_fraction}.

Of course, the choice of the search radius is always arbitrary, and varies in different studies from 100~mas to 1~arcsec
and more depending on the positional precision of compared catalogs and on the authors' judgment about the trade-off 
between the probability of a loss of an actual counterpart and the probability of the false identification.
\citet{Orosz2013} estimated the probability of the false identifications for matching of ICRF2 to SDSS. 
They found that the effect is significant for search radii larger than 500~mas.
Based on the data presented in Fig.~\ref{fig:cumulative_fraction}, it can be supposed that this probability
is small for OCARS, even when taking into account the higher accuracy of the {\gaia} positions as compared with SDSS.
Perhaps this is worth a further detailed investigation in a separate dedicated study following, e.g.,
the approaches proposed by \citet{Orosz2013} or \citet{Petrov2017a}.

Some cross-matching information is also included in the notes field in \verb"ocars.txt".
First, this field contains the ICRF source name if the source is included in the ICRF3.
Defining ICRF sources are marked with ``DEF''.
Other names in this field are primarily chosen to be recognized by both NED and SIMBAD.
The recognition is not always straightforward because both databases integrate data from many catalogs and often
the same physical object appears in different entries in NED and SIMBAD.
In addition, NED and SIMBAD have no unified rules for catalog codes.
In several cases, when the object is present in both databases but no common name is found, both names from NED
and SIMBAD are given in notes.
Several names can also be given when the same source is represented by several entries in NED or SIMBAD.
In any case, cross-identification included in \verb"ocars.txt" is incomplete, and the file \verb"ocars_n.txt" should
primarily be used as the cross-identification table between OCARS and other catalogs.

Some entries in VLBI-based position catalogs, especially based on a small number of observations,
may have large position errors or even be artifacts.
Therefore, the existence of radio source counterparts in different frequency bands can serve as a supplementary
proof of the reliability of the VLBI-derived position.

Table~\ref{tab:cross_identification} shows the number of OCARS sources cross-identified with other catalogs
in different frequency bands.
The catalog lists in the last column of this table are sorted in order of increasing priority of use when
the object names in several catalogs related to the same band are available.

\begin{table*}
\centering
\caption{Cross-identification between OCARS and other catalogs.}
\label{tab:cross_identification}
\begin{tabular}{lrl}
\hline
Catalog acronym  & Matched & \multicolumn{1}{c}{Data used} \\
Band             & sources & \multicolumn{1}{c}{(in order of increasing priority)} \\
\hline
NVSS                & 5696 & NRAO VLA Sky Survey \citep{Condon1998} \\
PMN                 & 2945 & Parkes-MIT-NRAO Survey \citep{Griffith1994} \\
87GB                & 3421 & Green Bank catalog 1987 \citep{Gregory1991} \\
SUMSS               &  769 & Sydney University Molonglo Sky Survey \citep{Mauch2003} \\
{\gaia}             & 4554 & {\gaia} Data Release 2 \citep{Lindegren2018,Mignard2018}\\
LQAC                & 5801 & Fourth LQAC release (LQAC-4) \citep{Gattano2018} \\
WISE/AllWISE        & 5328 & Wide-field Infrared Survey Explorer \citep{Wright2010,Mainzer2011} \\
Other IR            & 2024 & Infrared Astronomical Satellite mission (IRAS) \citep{Neugebauer1984} \\
                    &      & Two Micron All Sky Survey (2MASS) \citep{Skrutskie2006} \\
UV                  & 2261 & Galaxy Evolution Explorer (GALEX) \citep{Morrissey2007} \\
X-ray               & 1649 & ROSAT Survey \citep{White1994,Voges1999,Boller2016} \\
                    &      & Chandra X-Ray Observatory \citep{Evans2010} \\
                    &      & XMM-Newton Survey \citep{Watson2003,Watson2009,Rosen2016}\\
Gamma-ray           &  837 & AGILE sources \citep{Pittori2009} \\
                    &      & High Energy Stereoscopic System (HESS) \citep{Aharonian2005} \\
                    &      & Fermi Hard Gamma-ray LAT Catalog (FHL) \citep{Ackermann2013,Ackermann2016} \\
                    &      & Fermi Large Area Telescope Catalog (FGL) \citep{Abdo2010,Nolan2012,Acero2015} \\
\hline
\end{tabular}
\end{table*}


\section{Conclusion}
\label{sect:conclusion}

Knowledge of the physical characteristics of as large a number of radio sources as possible can play
an important role in addressing various astronomical and astrophysical problems.
The OCARS catalog supplements catalogs of radio source positions with data on object physical types, redshifts,
and optical and NIR photometry.
The bulk of this information for the OCARS catalog is taken from the NED and SIMBAD large astronomical databases,
which are supplemented with data found in the literature.
A principle of the work on the OCARS catalog is the need to update its contents immediately after the appearance
of new observational data.
On average, the catalog is updated every few weeks.

The OCARS catalog includes a separate file of photometric data in 14 visible and NIR bands.
Reliable data on the optical brightness of radio sources is required for the overall completeness
of the ICRF and other VLBI-based source positions catalogs.
Moreover, this data can also be used to address a variety of important tasks, such as the selection
of optically bright objects for radio-optical frame links, cross-identification of objects in different
catalogs, and separation of nearby objects in the same fields.
This last task is related to the fact that bright optical objects can be observed near the radio source,
and reliable identification of a particular optical object (or a component of a source) with the radio
source can prove a difficult problem.

Although the OCARS catalog is primarily intended to collect the information about the radio sources that have
accurate VLBI-derived positions, it also includes sources that were detected during VLBI experiments
but do not have yet a reliable position at a level of accuracy expected from astrometric VLBI.
These sources are included in the OCARS catalog to encourage their new VLBI observations, presumably on
more sensitive networks.

The OCARS catalog can be also useful for obtaining improved positions of the objects observed at UV, X-ray,
or gamma-ray bands, thereby offering new cross-identification suggestions in addition to those established
in the NED and SIMBAD.

By no means is the OCARS catalog intended to replace large general catalogs of extragalactic objects, such as
LQAC \citep{Gattano2018} or Milliquas \citep{Flesch2015}, which are essential for many studies.
The application of the OCARS catalog has a much narrower sphere than that of such comprehensive databases.
However, OCARS also has advantages, the most important of which are the use of a broad range of information sources,
active control of content by hand, and continuous maintenance.
Therefore, the OCARS catalog may be the most complete and accurate catalog of optical data for the astrometric radio
sources it contains.

The OCARS catalog is free for scientific research and other non-commercial use.
E-mail alerts about updates can be requested by interested users.

Finally, in spite of all the measures taken to verify the data included in the OCARS, the catalog is inevitably
not free of mistakes caused by possible incompleteness and inaccuracy of the data obtained from the literature,
and other astronomical catalogs and databases that serve as information sources for the OCARS catalog.
Possible errors in cross-identification of OCARS objects with other catalogs are another source of uncertainty.
Therefore, user feedback plays a very important role in correcting such errors and helping to further
develop the OCARS catalog.

\section*{Acknowledgments}

The author is grateful to Dmitry Blinov (St. Petersburg State University), Bryan Gaensler and Elizabeth Mahony
(University of Sydney), Anne-Marie Gontier and Jean Souchay (Paris Observatory), David Gordon and Karine Le Bail
(NASA Goddard Space Flight Center), Yuri Kovalev (Astro Space Center/Lebedev Physical Institute), Valeri Makarov
(U.S. Naval Observatory), Alexander Pushkarev (Crimea Astrophysical Observatory), Tapio Pursimo
(Nordic Optical Telescope), Michael Shaw (Stanford University), Marion Schmitz (NED team, IPAC, Caltech/JPL),
Jean Souchay and Fran\c{c}ois Taris (Paris Observatory), Oleg Titov (Geoscience Australia), and Vivienne Wild
(Institute of Astrophysics, Paris) for various help, fruitful discussions, and making their data available
for this work.

This work was partly supported by the Russian Government Program of Competitive Growth of Kazan Federal University.

This research has made use of the NASA/IPAC Extragalactic Database (NED), which is operated by the Jet Propulsion
Laboratory, California Institute of Technology, under contract with the National Aeronautics and Space Administration
\citep{Mazzarella2007}, SIMBAD database, operated at at the Centre de Donn\'ees Astronomiques (CDS), Strasbourg,
France \citep{Wenger2000}, and CfA-Arizona Space Telescope LEns Survey\footnote{https://www.cfa.harvard.edu/castles/}
(CASTLES).

Funding for the Sloan Digital Sky Survey IV has been provided by the Alfred P. Sloan Foundation,
the U.S. Department of Energy Office of Science, and the Participating Institutions. SDSS-IV acknowledges
support and resources from the Center for High-Performance Computing at the University of Utah.
The SDSS web site is www.sdss.org.

SDSS-IV is managed by the Astrophysical Research Consortium for the Participating Institutions of the SDSS
Collaboration including the Brazilian Participation Group, the Carnegie Institution for Science,
Carnegie Mellon University, the Chilean Participation Group, the French Participation Group,
Harvard-Smithsonian Center for Astrophysics, Instituto de Astrof\'isica de Canarias, The Johns Hopkins University,
Kavli Institute for the Physics and Mathematics of the Universe (IPMU) / University of Tokyo, Lawrence Berkeley
National Laboratory, Leibniz Institut f\"ur Astrophysik Potsdam (AIP), Max-Planck-Institut f\"ur Astronomie
(MPIA Heidelberg), Max-Planck-Institut f\"ur Astrophysik (MPA Garching), Max-Planck-Institut f\"ur Extraterrestrische
Physik (MPE), National Astronomical Observatories of China, New Mexico State University, New York University,
University of Notre Dame, Observat\'ario Nacional / MCTI, The Ohio State University, Pennsylvania State University,
Shanghai Astronomical Observatory, United Kingdom Participation Group, Universidad Nacional Aut\'onoma de M\'exico,
University of Arizona, University of Colorado Boulder, University of Oxford, University of Portsmouth,
University of Utah, University of Virginia, University of Washington, University of Wisconsin,
Vanderbilt University, and Yale University.

This work has made use of data from the European Space Agency (ESA) mission {\gaia} (https://www.cosmos.esa.int/gaia),
processed by the {\gaia} Data Processing and Analysis Consortium (DPAC, https://www.cosmos.esa.int/web/gaia/dpac/consortium)
\citep{Prusti2016,Brown2018}.
Funding for the DPAC has been provided by national institutions, in particular the institutions participating in the {\gaia}
Multilateral Agreement.

The author is grateful to the anonymous reviewer for careful reading of the manuscript and valuable comments and suggestions.

\bibliographystyle{aasjournal}
\bibliography{ocars}

\end{document}